\documentclass[submission,copyright,creativecommons]{eptcs/eptcs}
\providecommand{\event}{PLACES 2025}

\usepackage{iftex}

\ifpdf
  \usepackage{underscore}         
  \usepackage[T1]{fontenc}        
\else
  \usepackage{breakurl}           
\fi

\usepackage{listings}
\usepackage{multicol}
\usepackage{mathpartir}
\usepackage{stmaryrd}
\usepackage{textcomp}
\usepackage{pl-syntax}
\usepackage[numbers,square,sort&compress]{natbib}

\usepackage{xcolor}
\usepackage[normalem]{ulem}

\newcommand\codeidx[1]{\ensuremath{_\var{#1}}}
\newcommand\comm{\raisebox{0.5ex}{\texttildelow}>}
\newcommand\defcomm{define/<\raisebox{0.5ex}{\texttildelow}} 
\newcommand\sel{sel\comm}
\newcommand\branch{\rktil{branch?\ }}
\newcommand\chose{\rktil{choose!\ }}
\newcommand\var[1]{\texttt{#1}}
\newcommand\rktil[1]{\texttt{#1}}
\newcommand\project[2]{\ensuremath{\llbracket}\rktil{#1}\ensuremath{\rrbracket_{\var{#2}}}}

\title{Choreographies as Macros}
\author{Alexander Bohosian
\institute{Department of Computer Science and Engineering\\
  University at Buffalo, SUNY\\
  Buffalo, NY, USA}
\email{asbohosi@buffalo.edu}
\and
Andrew K. Hirsch
\institute{Department of Computer Science and Engineering\\
  University at Buffalo, SUNY\\
  Buffalo, NY, USA}
\email{akhirsch@buffalo.edu}
}
\newcommand{\titlerunning}{Choreographies as Macros}
\newcommand{\authorrunning}{A. Bohosian \& A. K. Hirsch}

\lstdefinestyle{racket}{
  basicstyle = \ttfamily,
  breaklines = true,
  escapeinside={\:}{\:},
  keepspaces
}

\begin{document}
\maketitle

\begin{abstract}
  Concurrent programming often entails meticulous pairing of sends and receives between participants to avoid deadlock.
Choreographic programming alleviates this burden by specifying the system as a single program.
However, there are more applications than implementations of choreographies, and developing new implementations takes a lot of time and effort.
Our work uses Racket to expedite building a new choreographic language called Choret.
Racket has a powerful macro system which allows Choret to reuse much of its infrastructure for greater functionality and correctness.


\end{abstract}

\section{Introduction}

Choreographic programming~\cite[see e.g.,][]{Montesi13,Montesi22,hirsch2022} is an emerging paradigm for designing and implementing concurrent systems.
Traditional languages require writing a program for each participant, while avoiding mismatched sends and receives that may cause deadlock.
This task grows more difficult with program size and the number of participants.
In contrast, choreographic programs---or \emph{choreographies}---encode the system's pattern of communication in a single program, rather than interweaving several programs.
This global view directly encodes the correct pairing of sends and receives, ensuring deadlock freedom.

Since choreographic programming is constantly evolving, rapid prototyping is often desirable.
Building a compiler from scratch---even a transpiler---takes time away from higher-level language decisions.
Ideally, one would reuse the existing infrastructure of a host language to provide a more robust implementation.
Recently, there has been significant interest in doing so via \emph{choreographic libraries}~\cite{KashiwaK24,ShenK24,ShenKK23,LugovicJ24}.
However, every current choreographic library either has nontraditional semantics or uses a nontraditional language design.
These nontraditional design choices allow the library to avoid issues arising from \emph{knowledge of choice}, a technical aspect of choreographic programming.
Ideally, we'd be able to implement the traditional design and semantics of choreographies in a choreographic library.
To do this, a clever \emph{scheme} is needed.

Such an implementation relies on the metaprogramming capabilities of the host language; in particular, the \textsc{Lisp} family of languages offer a powerful metaprogramming system in the form of macros.
In many \textsc{Lisp}s macros are user-defined functions which run during \emph{compilation}, specifically in a \emph{macro-expansion phase}.
These functions are written \emph{in \textsc{Lisp}}, akin to normal functions, and can perform arbitrary computation.
Most importantly, macros can be imported as part of a module, allowing language extensions to be used like normal libraries.

Intuitively, a choreographic language could be implemented as a set of \textsc{Lisp} macros, using their power to provide a choreographic library with traditional design and semantics.
We prove this intuition correct by building Choret, an embedded choreographic language in Racket.
Racket is a \textsc{Lisp} (specifically, an offshoot of Scheme) which offers a  particularly sophisticated macro system for metaprogramming, which is key for our implementation of knowledge of choice.

We provide the following contributions:
(a) in Section~\ref{sec:chor-chor}, we describe Choret via an example and describe its syntax;
(b) in Section~\ref{sec:epp-network-language} we describe the network (target) language of Choret, and describe Choret's compile-time semantics via \emph{Endpoint Projection}; and
(c) in Section~\ref{sec:rack-macro-expans} we describe how Racket's macro system allows us to implement Choret while maintaining traditional choreographic-language design and semantics.
Finally, in Section~\ref{sec:related-work} we survey related work, and in Section~\ref{sec:conclusion} we conclude.


\section{Choreographies and Choret}
\label{sec:chor-chor}

Let us begin by considering an example concurrent application, and see how we would implement it both as a traditional Racket program and in Choret.
We use a traditional example: the bookseller.
Here, a \rktil{Buyer} wants to buy a book from a \rktil{Seller}.
To do so, \rktil{Buyer} sends \rktil{Seller} the title of the book and then \rktil{Seller} looks up the title in a catalog and sends back a price.
\rktil{Buyer} then determines whether the price is within their budget.
If it is, then \rktil{Buyer} informs \rktil{Seller} of this choice and sends their address, and \rktil{Seller} sends \rktil{Buyer} a date by which to expect the book.
If the book is not within \rktil{Buyer}'s budget, then they inform the \rktil{Seller} of this and the protocol ends.

Bellow, we can see a Racket program which implements this protocol in the traditional style.
There is one program for each of the \rktil{Buyer} and the \rktil{Seller}.
These programs share a channel~\rktil{ch}, which they use to send and receive messages to and from each other.
The sends and receives need to be perfectly matched, and this needs to be done by hand: the programmer must check to make sure that they are following the protocol precisely.
We have made it easier to read our example by adding subscripts to the sends and receives: $\rktil{send}_n$ matches with $\rktil{recv}_n$.
\begin{multicols}{2}[]
  \begin{lstlisting}[style = racket]
;; Code at Buyer
(send:\codeidx{1}: ch title)
(recv:\codeidx{2}: ch cost)
(if (<= cost budget)
  (block
    (send:\codeidx{3}: ch "buy")
    (send:\codeidx{4a}: ch address)
    (define date (recv:\codeidx{5a}: ch)))
  (block
    (send:\codeidx{3}: ch "nevermind")
    (define response (recv:\codeidx{4b}: ch))))
  \end{lstlisting}
  \columnbreak
  \begin{lstlisting}[style = racket]
;; Code at Seller
(define title (recv:\codeidx{1}: ch))
(send:\codeidx{2}: ch (catalog title))
(define response (recv:\codeidx{3}: ch))
(if (eq? response "buy")
  (block
    (define address (recv:\codeidx{4a}: ch))
    (send:\codeidx{5a}: ch (ship title address))
  (block
    (send:\codeidx{4b}: ch "goodbye")))
  \end{lstlisting}
\end{multicols}

Note the painstaking way that the sends and receives need to be matched.
In particular, note that there are two copies of $\rktil{send}_3$, each matched with the same $\rktil{recv}_3$: because they are in different branches of an \rktil{if} expression, only one of these \rktil{send}s will run.
However, it further complicates the job of matching these sends and receives.
While this matching is easily possible to perform in this case, in larger programs it can become very difficult.

This is where Choret (and choreographic programming more broadly) comes in.
Rather than writing a program for each of the participants, in Choret we would instead write one canonical program for the entire system and then compile that single program into the two separate programs above.
By doing so, we match sends and receives syntactically, making the matching automatic.
Thus, we would write \rktil{(\defcomm{} (at $P$ $x$) (at $Q$ $e$))} to say ``evaluate the expression~$e$ on process~$Q$, and then send the resulting value to $P$, where it should be called~$x$.''
Here, \rktil{(at $Q$ $e$)} just means ``evaluate the expression~$e$ on process~$Q$.''

There is another kind of communication that occurs in Choret: communication that is used to propagate \emph{knowledge of choice}.
We can see this above: when the \rktil{Buyer} determines whether the cost of the book is within their budget, the \rktil{Seller} has no idea which choice the \rktil{Buyer} made.
Instead, we see the \rktil{Buyer} informing the \rktil{Seller} of their choice using the two copies of $\rktil{send}_3$.
The seller then branches on this by asking if the result of $\rktil{recv}_3$ is \rktil{"buy"}, and behaving appropriately in each case.
The Choret program \rktil{(\sel\ $P$\ ([$l$\ $Q$]) $E$)} means ``$P$ informs $Q$ that they are taking the branch labeled $l$, and then the entire system continues as $E$.''
Combining this with the basic communication primitives above, we can rewrite the example above into a single Choret program:
\begin{lstlisting}[style = racket]
(chor (S B)
  (define/<~:\codeidx{1}: (at S title) (at B title))
  (define/<~:\codeidx{2}: (at B cost) (at S (catalog title))
  (if (at B (<= cost budget))
    (sel~>:\codeidx{3}: B ([S 'buy])
      (define/<~:\codeidx{4a}: (at S address) (at B address))
      (define/<~:\codeidx{5a}: (at B date) (at S (ship title address))
    (sel~>:\codeidx{3}: B ([S 'do-not-buy])
      (define/<~:\codeidx{4b}: (at B response) (at S "goodbye"))))))
\end{lstlisting}

Not only is the choreography shorter and more concise, but it's no longer possible to mismatch the pairs of sends and receives.
Because of this, choreographies offer an exciting property: deadlock freedom by construction~\cite{carbone2013}.
Thus, users of Choret can write their code without fear of deadlocks.

\begin{figure}
  \centering
  \begin{syntax}
    \abstractCategory[Racket Expressions]{e}
    \category[Binding Forms]{B} \alternative{X} \alternative{\rktil{(at $P$ $x$)}}
    \category[Choret Programs]{P} \rktil{(chor ($P$ \dots) $T$ \dots)}
    \category[Choret Expressions]{E}
      \alternative{\rktil{(at $P$ $e$ \ldots)}}
      \alternative{\rktil{(\comm{} (at $P$ $e$) $Q$)}}\\
      \alternative{\rktil{(if (at $P$ $e$) $E_1$ $E_2$)}}
      \alternative{\rktil{(\sel{} $P$ ([$l$ $Q$] \dots) $E$)}}\\
      \alternative{\rktil{(let ([$B$ $E_1$] \dots) $E$)}}
      \alternative{\rktil{(let* ([$B$ $E_1$] \dots) $E$)}}\\
      \alternative{\rktil{(set! (at $P$ $x$) $E$)}}
    \category[Choret Terms]{T}
      \alternative{\rktil{(define $B$ $E$)}}
      \alternative{\rktil{(\defcomm{} (at $P$ $x$) (at $Q$ $e$))}}
    \end{syntax}
  
  \caption{Choret Syntax}
  \label{fig:choret-syn}
\end{figure}

Formally, the syntax of Choret is given in Figure~\ref{fig:choret-syn}.
Like Racket more generally, Choret is split into terms (i.e., top-level definitions) and expressions (which return a value).
Definitions, whether top-level or local, can either bind global variables or local variables at some process~$P$.
For instance, the Choret term \rktil{(\defcomm{} (at $P$ $x$) (at $Q$ $e$))} binds the local variable~$x$ at process~$P$.

By contrast, \rktil{(at $P$ $e$)} and \rktil{(\sel{} $P$ ([$l$ $Q$]) $E$)} are both expressions.
Selection is more powerful than previously suggested: $P$ can send any number of labels to (distinct) processes, informing them all of the taken branch.
The  expression \rktil{(\comm{} (at $P$ $e$) $Q$)} computes the value of $e$ at $P$ and then sends the result to $Q$.
Finally, we include traditional Racket expressions as Choret expressions, often extending them to describe where computation is taking place via the \rktil{at} syntax.

Our treatment of knowledge of choice via selection is the tradition in choreographic programming-language design~\cite{Montesi13,hirsch2022,Montesi22}.
However, it leads to significant difficulties in developing choreographic libraries.
As we will see in Section~\ref{sec:epp-network-language}, the use of selections means that the process of splitting a choreography into different programs for each process requires multiple passes.
However, most metaprogramming systems do not make writing multipass transformations possible, much less easy.
Therefore, most choreographic libraries either change how they handle knowledge of choice or they perform the splitting at runtime.
Either choice allows them to avoid the multiple passes required at compile time.
Racket's macro system allows us to uniquely provide the traditional design with its traditional semantics (see Section~\ref{sec:rack-macro-expans}).


\section{EPP and the Network Language}
\label{sec:epp-network-language}

Choreographies give a global view of the system.
However, in order to execute a choreography, we must spit it into separate programs, one for each participant.
In the choreographic literature, this transformation is referred to as \emph{Endpoint Projection~(EPP)}.
In this section, we describe the design of our network language (that is, the target of endpoint projection) as well as the definition of endpoint projection itself.
We discuss their Racket implementation in Section~\ref{sec:rack-macro-expans}.

\begin{figure}
  \centering
    \begin{syntax}
      \category[Network]{N}
      \alternative{\ldots \textit{(All other Racket forms)}}\\
      \alternative{\rktil{(send $P$ $e$)}}
      \alternative{\rktil{(recv $P$)}}\\
      \alternative{\rktil{(\chose $P$ $l$ $N$)}}
      \alternative{\rktil{(\branch $P$ ([$l$ $N$] \dots))}}
    \end{syntax}
  \caption{Network Language Syntax}
  \label{fig:net-syn}
\end{figure}

The syntax of our network language can be found in Figure~\ref{fig:net-syn}.
Unlike Choret itself, which manually reimplements the core Racket forms, our network language is described as four additions on top of Racket itself (implemented as simple macros).
The form~\rktil{(send $P$ $e$)} evaluates the Racket expression~$e$ and sends the result to the process~$P$, which is assumed to be different from the current process.
Similarly, the form~\rktil{(recv $P$)} receives a value from the process~$P$, returning that value.
We propagate knowledge of choice via the forms \rktil{(\chose $P$ $l$ $E$)} and \rktil{(\branch $P$ ([$l$ $E$] \dots))}.
The former informs $P$ about which branch was taken, while the latter allows $P$ to tell the current process which branch to take.

We now almost have enough information to formally define EPP.
However, one difficulty arises, which is best described by example.
Consider the following Choret program:
\begin{lstlisting}[style=racket]
  (chor (A B)
    (define (at A x) ...)
    (if (at A x)
      (sel~> A [B l]
        (at B "Left"))
      (sel~> A [B r]
        (at B "Right"))))
\end{lstlisting}
Here, we look at $A$'s boolean value and, if it's true, $B$ returns \rktil{"Left"}, otherwise $B$ returns \rktil{"Right"}.
In order to allow this difference in $B$'s behavior, $A$ informs $B$ of which branch to take.
Now, imagine trying to project a program for $B$ from this choreography.
Projecting each branch of the \rktil{if} is easy: the true branch projects to \rktil{(\branch $A$ [l "Left"])}, the false to \rktil{(\branch $A$ [r "Right"])}.
These each wait for a message from $A$ and either return \rktil{"Left"} if the message is \rktil{l} (for the former) or return \rktil{"Right"} if the message is \rktil{r} (for the latter), doing nothing otherwise.
In order to give a single program for $B$, we need a program which receives a message from $A$ and takes both behaviors, doing nothing only if the message is neither \rktil{l} nor \rktil{r}.
We do this via \emph{merging}.
We define merging, written $N_1 \sqcup N_2$, as follows:
$$
N_1 \sqcup N_2 = \left\{
  \begin{array}{ll}
    \text{recursively merge} & \text{if $N_1$ and $N_2$ are matching Racket forms}\\
    \rktil{(send $P$ $e$)} & \text{if $N_1 = N_2 = \rktil{(send $P$ $e$)}$}\\
    \rktil{(recv $P$)} & \text{if $N_1$ = $N_2$ = \rktil{(recv $P$)}}\\
    \rktil{(\chose $P$ $l$ $N_1' \sqcup N_2'$)} & \begin{array}[t]{l}\text{if $N_1 = \rktil{(\chose $P$ $l$ $N_1'$)}$}\\ \text{and $N_2 = \rktil{(\chose $P$ $l$ $N_2'$)}$}\end{array}\\
    \rktil{(\branch $P$ $(\begin{array}[t]{l}[l_{1i}\ N_{1i} \sqcup N_{2j}] \dots\\{} [l_{1k}\ N_{1k}] \dots\\{} [l_{2k}\ N_{2k}]))\end{array}$} & \begin{array}[t]{l}\text{if $N_1 = \rktil{(\branch $P$ ([$l_{11}$\ $N_{11}$] \ldots))}$}\\ \text{and $N_2 = \rktil{(\branch $P$ ([$l_{21}$\ $N_{21}$] \ldots))}$}\\ \text{and $l_{1i} = l_{2j}$}\\ \text{and}~\forall k, k'. l_{1k} \neq l_{2k'}\end{array}\\
    \bot & \text{otherwise}
  \end{array}
\right.
$$

The merge function looks intimidating, but the only complicated case is \rktil{branch}; every other case merely checks to make sure that $N_1$ and $N_2$ are compatible before making a recursive call.
In the case of \rktil{branch}, we need to combine the possible branches.
If both $N_1$ and $N_2$ have a branch for some label $l$, then we recursively call merge on those branches.
Any labels that either $N_1$ or $N_2$ have, but not both, are simply kept.
If we apply this to our example above, we compute
$$\begin{array}{c}\rktil{(\branch $A$ ([l "Left"]))} \sqcup \rktil{(\branch $A$ ([r "Right"]))} =\\ \rktil{(\branch $A$ ([l "Left"] [r "Right"]))}\end{array}$$
which behaves exactly as desired.

\begin{figure}
  \centering
  {\small
    $$
    \project{E}{A} = \left\{
      \begin{array}{ll}
        e\ \ldots                        & \text{if}~E = \rktil{(at $A$ $e$ \ldots)}\\
        \rktil{(void)}                               & \text{if}~E = \rktil{(at $P$ $e$ ...)} \mathrel{\text{where}} P \neq A\\
        \rktil{(send $Q$ $e$)}          & \text{if}~E = \rktil{(\comm\ (at $A$ $e$) $Q$)}\\
        \rktil{(recv $P$)}              & \text{if}~E = \rktil{(\comm\ (at $P$ $e$) $A$)}\\
        \rktil{(void)} &  \text{if}~E = \rktil{(\comm\ (at $P$ $e$) $Q$)} \mathrel{\text{where}} P \neq A \mathrel{\text{and}} Q \neq A\\
        \rktil{(if $e$ \project{$E_1$}{A} \project{$E_2$}{A})} & \text{if}~E = \rktil{(if (at $A$ $e$) $E_1$ $E_2$)}\\
        \rktil{$\project{$E_1$}{A}\sqcup \project{$E_2$}{A}$} & \text{if}~E = \rktil{(if (at $P$ $e$) $E_1$ $E_2$)} \mathrel{\text{where}} P \neq A\\
        \rktil{(let ([$X$ \project{$E_1$}{A}] \ldots) \project{$E$}{A})} & \text{if}~E = \rktil{(let ([$X$ $E_1$] \ldots) $E$)}\\
        \rktil{(let ([$x$ \project{$E_1$}{A}] \ldots) \project{$E$}{A})} & \text{if}~E = \rktil{(let ([(at $A$ $x$) $E_1$] \ldots) $E$)}\\
        \rktil{(let ([$\_$ \project{$E_1$}{A}] \ldots) \project{$E$}{A})} & \text{if}~E =
                                                                       \begin{array}[t]{l}
                                                                         \rktil{(let ([(at $P$ $x$) $E_1$] \ldots) $E$)}\\
                                                                         \mathrel{\text{where}} P \neq A\\
                                                                       \end{array}\\
        \begin{array}[t]{l}
          \rktil{(\chose $Q_1$ $l_1$ }\\
          \rktil{    \project{(\sel\ $A$ ([$l_2$ $Q_2$] \ldots) $E$)}{A})}\\
        \end{array} & \text{if}~E =
                      \begin{array}[t]{l}
                        \rktil{(\sel\ $A$ ([$l_1$ $Q_1$] [$l_2$ $Q_2$] \ldots) $E$)}\\
                      \end{array}\\
        \begin{array}[t]{l}
          \rktil{(\branch $P$ }\\
          \rktil{    ([$l_1$ \project{(\sel\ $P$ ([$l_2$ $Q_2$] \ldots) $E$)}{A}]))}\\
        \end{array} & \text{if}~E =
                      \begin{array}[t]{l}
                        \rktil{(\sel\ $P$ ([$l_1$ $A$] [$l_2$ $Q_2$] \ldots) $E$)}\\
                      \end{array}\\
        \project{(\sel\ $P$ ([$l_2$ $Q_2$] \ldots) $E$)}{A} & \text{if}~E =
                                                         \begin{array}[t]{l}
                                                           \rktil{(\sel\ $P$ ([$l_1$ $Q_1$] [$l_2$ $Q_2$] \ldots) $E$)}\\
                                                           \text{where}~P\neq A \mathrel{\text{and}} Q_1 \neq A
                                                         \end{array}\\
      \end{array}
    \right.
    $$
  }
  \caption{Definition of Endpoint Projection (Selected Parts)}
  \label{fig:epp}
\end{figure}

We use the definition of merging to define endpoint projection in Figure~\ref{fig:epp}.
This describes how to transform each of the forms of Choret into a network-language form.
As an example, a choreographic send \rktil{(\comm\ (at $P$ $e$) $Q$)} is transformed into \rktil{(send $Q$ $e$)} for~$P$ and into \rktil{(recv $P$)} for~$Q$.
For any process not involved, a Choret form will turn into \rktil{(void)}, a Racket standard-library function which returns ``nothing.''
Thus, every process only gets the information available to them in the choreography.

Note that Figure~\ref{fig:epp} only contains selected forms.
The other forms are uninteresting; they recursively call EPP on their subforms and then return an ``obvious'' Racket analog of themselves.
This pattern can be seen in the \rktil{let} lines of Figure~\ref{fig:epp}.

Now that we have a mathematical definition of EPP, we can begin to implement it in Racket macros.
However, this leads to some complications: Racket's macro system is powerful, but expressing complicated, multi-pass transformations like EPP inside of them is still difficult.
In Section~\ref{sec:rack-macro-expans}, we introduce the tricks and tips that the Racket community has put together for this problem and describe how we use those tricks to implement Choret.


\section{Racket Macro Expansion and EPP}
\label{sec:rack-macro-expans}

We now explain the implementation of Choret.
This implementation relies on the power of Racket's macro system to allow us to perform merging and EPP at compile time.
Thus, we begin by providing some background on the Racket macro system before finally describing the implementation of EPP using that system.

\subsection{Background on Racket Macros}
\label{sec:rack-macro-explanation}

Racket canonizes its own syntax into data called \emph{syntax objects}.
These syntax objects are Racket data that represent Racket programs; \emph{macros} are then simply functions that take and return syntax objects.
Syntax objects themselves contain not only the abstract-syntax tree of a program, but also information such as scope and source locations.
Thus, macros form a powerful metaprogramming facility.

Racket's macro expansion is performed top-down, outermost to innermost, and fully expands all macros to their \emph{core forms}.
Thus, a macro ``sees'' its arguments in unexpanded form.
While this is normally desirable, sometimes a programmer wants a macro to operate on \emph{expanded} output.
In order to do this, the macro calls \rktil{local-expand}, which invokes the macro expander directly.
Calling \rktil{local-expand} on a syntax object returns its full expansion, which can then be parsed and analyzed using Racket's \rktil{syntax-case} form.

The Racket core form \rktil{quote-syntax} turns data into a syntax object.
Syntax objects obtained this way retain most of their lexical information (i.e. scope sets)~\cite{quote-syntax-racket25}, though certain scopes are pruned.
However, a programmer can force \rktil{quote-syntax} to preserve all scopes using the \rktil{\#:local} keyword.
Since \rktil{quote-syntax} is defined as a core form, the macro expander will not touch it or the data it is transforming.
Thus, programmers can use \rktil{quote-syntax} to prevent the macro expander from expanding some syntax, preserving it for other macros to see.

Finally, sometimes a programmer wants to communicate information across macros nonlocally; for instance, they may want to allow the expansion of one macro to determine how another expands globally.
To do so, they can store that information in a \emph{syntax parameter}.
Syntax parameters allow for dynamic macro time bindings,
which can be used to update a binding for expansions within an entire branch of the syntax tree.

\subsection{The Implementation of EPP}
\label{sec:epp-explanation}

In order to implement select-and-merge EPP as a library during compile time, we take full advantage of the Racket macro system.
The top-level \rktil{chor} macro creates a syntax parameter representing the process currently being expanded.
It then loops through all of the processes in the choreography, expanding its body once for each process, setting its syntax parameter appropriately each time.

Most other macros in Choret don't use \rktil{local-expand} and instead directly rearrange them according to the EPP specification from Section~\ref{sec:epp-network-language}, implicitly relying on the macro expander to expand their subforms.
As described in that specification, they produce programs in our network language, which is also a collection of macros.

For most of Choret's expressions, this works beautifully.
However, selections create \rktil{branch}es, which need to be merged later.
If we allowed the \rktil{branch} macro to be expanded fully to its core forms, the \rktil{merge} macro would not be able to detect when two branches need to be merged.
Thus \rktil{(\branch $P$ ([$l$ $E$]))} expands to another \branch form wrapped with \rktil{quote-syntax}, with the $E$ subform expanded, like \rktil{(quote-syntax (\branch $P$ ([$l$ \project{$E$}{A}])) \#:local)}.
The \rktil{merge} macro can then look for these hidden \rktil{branch}es without fear that they will be expanded away by Racket.

All together, this design allows us to provide a traditional select-and-merge choreographic language design without requiring EPP to be performed at runtime.
Other choreographic libraries, such as HasChor~\cite{ShenKK23}, mix the semantics of their network languages with EPP, allowing them to project the appropriate branch when required rather than performing full merges.
This means two things.
First, every process sees the entire choreography, which may not be appropriate in mixed-trust settings.
Second, EPP can now block the execution of a program, potentially slowing down a system considerably.
However, Racket's type system enables Choret to perform EPP fully at compile time.


\section{Related Work}
\label{sec:related-work}

\subsection{Choreographic Programming}
\label{sec:chor-progr}

Choreographic programming emerged from the process-calculus and session-type communities about ten years ago~\cite{carbone2012,carbone2013,Montesi13,Montesi22}.
Since that time, most of the work has been exploring theoretical aspects of EPP in lower-order settings: there was no ability to create subroutines or functions~\cite{carbone2013,montesi2015,cruz-filipe2016a,cruz-filipe2016,cruz-filipe2017c,cruz-filipe2017a,giallorenzo2018}.
Recently a fair amount of interest has come up in \emph{functional} choreographic programming, which combines choreographic programming languages with $\lambda$~calculi to allow for program abstraction and reuse~\cite{hirsch2022,cruz-filipe2022,GraversenHM24}.
In particular, this work is inspired by Pirouette, the first functional choreographic programming language~\cite{hirsch2022}.

As an outgrowth on the work on functional choreographic programming, many people have begun to experiment with \emph{embedded} choreographic languages.
This was started by HasChor~\cite{ShenKK23}, which embedded a choreographic programming language inside of Haskell using a freer monad.
This method made it easy to implement a choreographic language.
However, it lead to an unusual situation: endpoint projection was no longer a compile-time activity, but something that happened at runtime whenever a node needed the next line of its instructions.
This methodology, with slight variations, has since been adapted by other embedded implementations of choreographic languages~\cite{KashiwaK24,ShenK24}.

The closest work to this is Klor, which is an embedded implementation of choreographies in Clojure, another Scheme-like language~\cite{LugovicJ24}.
Like our work, Klor uses macros to implement a choreographic language inside of a LISP-like language with EPP at compile time.
However, Klor handles knowledge of choice significantly differently.
Whereas we, like most of the choreographic literature~\cite{Montesi13,Montesi22,GraversenHM24,cruz-filipe2020,cruz-filipe2022,hirsch2022}, use selection messages to encode knowledge of choice, Klor instead uses a relatively new idea: agreement types~\cite{BatesN24}.
These allow any data to be located at a \emph{collection} of processes, and communication adds a process to that collection.
When a choreography branches on data, then, any process in that collection knows which path to take.
Doing so allows Klor to avoid merging, and therefore avoid the need for \rktil{local-expand}.
We, in contrast, choose to implement the traditional approach to knowledge of choice in choreographies.
We are thus the only embedded implementation of choreographies with traditional select-and-merge EPP at compile time.

\subsection{Embedded Languages via Racket Macros}
\label{sec:embedd-lang-rkt}

The defining feature of Racket is its ``Languages as Libraries'' design,
which entails an API for extending the language.
A good example is Typed Racket, a sister language of Racket implemented entirely as a normal Racket library~\cite{Tobin-Hochstadt2008,Tobin-Hochstadt2011}.

A characteristic feature of many \textsc{Lisp}s, like Racket, is the ability to easily embed domain specific languages (DSLs) using macros.
However, macros in other \textsc{Lisp}s tend to have certain issues.
For example, macros in many \textsc{Lisp}s use \emph{symbols}---essentially immutable strings---to encode identifiers.
However, this allows macros to ignore scope when manipulating identifies: a macro may introduce identifiers which accidentally capture identifiers in the macro's body, or it may introduce identifiers which are accidentally captured by bindings in the macro's body.
Such problems, among others, are often referred to as \emph{macro hygiene}.
Racket largely solves such issues using scope sets~\cite{Matthew2016},
which associates with each identifier a set of scopes.
Racket uses such scope sets to determine the correct bindings for the expanded code.

While scope sets ensure macro hygiene, it is sometimes desirable to use unhygienic macros.
For example, Choret sometimes needs to communicate which participant is currently being projected to the macros that perform EPP.
We are able to do this by using Racket's \rktil{syntax-parameter}~\cite{Barzilay2011} macro,
which, when expanded, updates a compile time binding that is only visible in the body of the macro.
Thus, we are able to selectively ignore hygiene when necessary, while writing hygenic macros by default.


\section{Conclusion}
\label{sec:conclusion}

Choreographies are a promising paradigm for concurrent programming.
However, in order for them to live up to their promise, the community needs to rapidly develop and prototype new choreographic-language designs.
Choreographic libraries are a promising method for doing so, but they rely on the metaprogramming capabilities of a host language.
Because these capabilities tend to be weak, previous choreographic-library designers have developed clever new semantics for choreographic languages which can be implemented with those anemic capabilities.
However, that has left the most-common design for choreographic languages---the traditional select-and-merge semantics---without the rapid prototyping advantages of choreographic libraries.

By developing Choret, we have shown that Racket's macro system is strong enough to bridge this gap.
In particular, we have implemented Choret as a choreographic library in Racket that performs EPP at compile time.
While Choret uses advanced features of Racket's macro system, the implementation is quite small---only 370 lines of code (excluding comments and tests).
We hope that this small size means that other choreographic-language designers will build on Choret in order to test out their designs.


\bibliographystyle{eptcs/eptcs}
\bibliography{main}

\end{document}